\documentclass[secnumarabic,aps,pra,amsfonts,twocolumn,%
showkeys,floatfix]{revtex4}

\usepackage{bm,graphicx}

\newcommand{\gl}[1]{(\ref{#1})}

\begin{document}

\title{Modified BCS mechanism of Cooper pair formation in narrow  energy
  bands of special symmetry\\III. Physical interpretation}
\author{Ekkehard Kr\"uger}
\email{krueger@physix.mpi-stuttgart.mpg.de}
\affiliation{Max-Planck-Institut f\"ur Metallforschung, D-70506 Stuttgart,
  Germany}
\date{\today}
\begin{abstract}
  In Part I of this paper a modified BCS mechanism of Cooper pair formation
  was proposed. The present Part III gives a physical
  interpretation of this mechanism in terms of spin-flipping
  processes in superconducting bands.
\end{abstract}
\maketitle
\section{Introduction}
\label{introduction}
In Part I of this paper \cite{josn} a modified BCS mechanism of Cooper pair
formation was proposed that operates in narrow ``superconducting'' energy
bands ($\sigma$ bands) and can be established within a nonadiabatic
extension of the Heisenberg model of magnetism \cite{hei}. This
``nonadiabatic Heisenberg model'' (NHM) is defined by three postulates given,
e.g., in \cite{enhm}.

The present paper gives a physical interpretation of this mechanism. In
Section \ref{sec:1} the special properties of the spin-dependent Wannier
functions (spin-dependent Wfs) in narrow $\sigma$ bands are interpreted in
terms of spin-flipping processes. In Sections \ref{sec:2} and \ref{sec:3} the
spin-phonon coupling in narrow $\sigma$ bands and the resulting mechanism of
Cooper pair formation, respectively, are illustrated in terms of these spin
processes.

In Section \ref{sec:4} the important difference between the new mechanism in
narrow $\sigma$ bands and the conventional BCS mechanism \cite{bcs} is
interpreted in terms of ``spring-mounted'' Cooper pairs.

\section{Spin-flipping processes}
\label{sec:1}
Assume a metal to be given that possesses a narrow, roughly half-filled
$\sigma$-band complex in its calculated band structure. As in Part I we
assume that this metal has only one atom in the unit cell. That means that
the $\sigma$-band complex consists only of a single $\sigma$ band.

The Bloch functions of the $\sigma$ band can be transformed into
spin-dependent Wfs $w_m(\vec r - \vec T, t)$ [as given in Eq.~(A14) of Part
I] with the following properties:

(i) they are centered on the atomic
positions $\vec T$;

(ii) they are obtained by a {\em unitary}
  transformation from the Bloch functions of the $\sigma$ band; 
  
(iii) they are adapted to the symmetry of the considered metal; and
  
(iv) they are as well localized as possible.

\begin{figure}
\includegraphics[width=.4\textwidth]{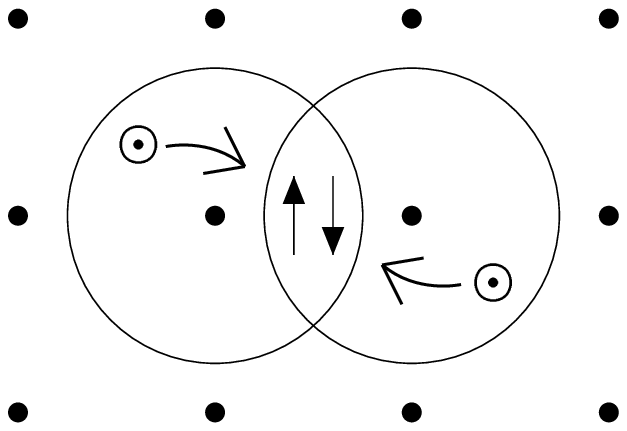}
\caption{ Spin-flipping process in a narrow $\sigma$ band: two electrons with
  parallel spins (represented by small circles $\odot$) in adjacent
  localized states comply with the Pauli principle by aligning their spins
  antiparallel when they approach.}
\label{figure1}
\end{figure}

\noindent
($m = \pm \frac{1}{2}$ is the quantum number of the crystal spin; $\vec r$
and $t$ stand for the local and spin coordinate, respectively, of the
electron.)  However, the spin-dependent Wfs cannot be written as a product of
a local function $w(\vec r - \vec T)$ with the spin functions $u_s(t)$ ($s$
denotes the spin quantum number) because the matrix $[f_{sm}(\vec k)]$ in
Eq.~(A9) of \mbox{Part I} cannot be chosen independent of $\vec k$ in a
$\sigma$ band.  Thus, these functions represent localized states $|\vec
T,m\rangle$ with spin directions depending on the positions $\vec r$ of the
electrons.  This motion may be described in terms of ``spin-flipping
processes'' as, e.g., illustrated in Fig.~\ref{figure1}. Here two electrons
with parallel spins align their spins antiparallel when they approach within
the region of overlap.

It is essential that within the NHM the states $|\vec T,m\rangle$ are {\em
  really occupied by electrons}. This is a consequence of the second
postulate of the NHM, i.e., of the special electron-electron interaction in
the nonadiabatic system which is forced by this postulate. Hence, within the
NHM the spin-flipping processes in a narrow $\sigma$ band are produced by a
{\em real electron-electron interaction} that operates in the nonadiabatic
system.

The flippings of the spins are a consequence of the special mathematical
properties of the spin-dependent Wfs $w_m(\vec r - \vec T, t)$ which are
determined by the four above conditions (i) - (iv) that must be satisfied
within the NHM. Hence, we understand the physical character of this
interaction if we understand why the $w_m(\vec r - \vec T, t)$ must comply
with these conditions.

The first and the fourth condition (i) and (iv) are required by the second
postulate. Hence, these conditions allow the electrons to take the state of
lowest Coulomb repulsion energy. The second condition (ii) yields a
nonadiabatic Hamiltonian with the correct symmetry and the third condition
(iii) guarantees that the spin-dependent Wfs are orthogonal. Hence, the third
condition is required that electrons occupying the states $|\vec T,m\rangle$
comply with the Pauli principle.

Consequently, in the ground state of a narrow $\sigma$ band {\em the
  electrons change permanently their spin directions in order to comply with
  the Pauli principle.}

\section{Spin-phonon interaction}
\label{sec:2}
In a narrow $\sigma$ band, the operator $H_{\rm Cb}^{n\sigma}$ of Coulomb
interaction has the rather complicated form given in Eq.~(3.13) of Part I.
$H_{\rm Cb}^{n\sigma}$ depends on boson as well as on fermion operators and
represents a special ``spin-phonon interaction'' which may be understood as
follows.

\begin{figure}
\includegraphics[width=.4\textwidth]{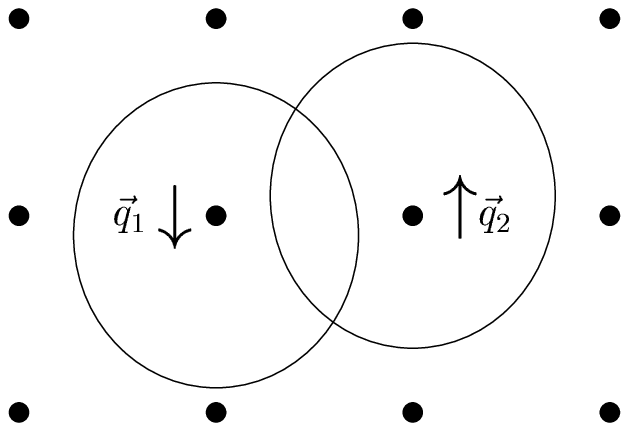}
\caption{ The spin-flipping process indicated in Fig.~\ref{figure1} produces
  asymmetric charge distributions that accelerate the positive ions in such a
  way that a phonon pair is emitted or absorbed. $\vec q$ stands for the
  acceleration of the positive ions.}
\label{figure2}
\end{figure}

On a spin-flipping process as indicated in Fig.~\ref{figure1} the spin
angular momentum of the two electrons is not conserved. During this process
the electrons behave like gyroscopes: inertial forces act
on the electrons that deform their orbitals within their localized states. As
indicated in Fig.~\ref{figure2}, the localized charge distributions become
asymmetric with respect to the lattice and accelerate the positive ions in
such a way that phonons are emitted or absorbed.  Eq.~(3.14) of Part I shows
that $H_{\rm Cb}^{n\sigma}$ conserves the total crystal spin angular
momentum.

\section{Cooper pair formation}
\label{sec:3}
As a consequence of the spin-phonon coupling in a narrow $\sigma$ band,
crystal-spin-1 phonons are excited in the ground state of the nonadiabatic
system.  We may assume that at zero temperature these phonons are virtually
excited, i.e., each phonon pair is reabsorbed immediately after its
generation.  Therefore, in Part I we approximated the nonadiabatic
Hamiltonian by a purely electronic Hamiltonian $H^0$ in order to describe the
electron-phonon system at zero temperature; see Eq.~(3.17) of Part I.

$H^0$ satisfies the equation of constraint (3.36) in \mbox{Part I}. This
equation distinguishes the mechanism of Cooper pair formation within the NHM
from the conventional BCS mechanism. As a consequence of this equation of
constraint, in {\em all the eigenstates} of $H^0$ the electrons form Cooper
pairs. This may be understood as follows.

The system represented by $H^0$ is purely electronic though the spin-flipping
processes still occur. The electron spins still couple to the motion of the
atomic cores and, hence, to the phonons. These phonons, however, {\em cannot
  store} crystal spin angular momenta but must be reabsorbed immediately
after their generation. In this way, the phonons mediate a special spin-spin
interaction between the electrons.

Thus, we have to look for a state with the total spin angular momentum being
conserved within the electron system {\em alone}. That means, within the
system represented by $H^0$ every spin-flipping process of two $\sigma$-band
electrons must be accompanied by the opposite flipping process of two other
$\sigma$-band electrons. This opposite process may occur somewhere in the
crystal since the electron spins interact via virtual crystal-spin-1
phonons.

In this context, ``opposite'' means ``time inverted'' because the
time-inversion symmetry is the only symmetry which makes such an opposite
motion possible.  Hence, the total spin angular momentum is conserved within
the electron system if all the electrons form pairs that are invariant under
time inversion. That means, whenever a certain one-electron state is
occupied, then its time-inverted state must be occupied, too. In other words,
all the electrons must form Cooper pairs.

\section{Constraining forces}
\label{sec:4}

We obtain further information on the spin-flipping processes in a narrow
$\sigma$ band by transforming the operator $H_{\rm Cb}^{n\sigma}$ of Coulomb
interaction into the
$\vec k$ representation,
\begin{widetext}
\begin{equation}
\label{eq:2}
H_{\rm Cb}^{n\sigma} = \displaystyle\sum_{\vec k, m}
\langle\vec k_{1}'', l_{1}; 
\vec k_{2}'',l_{2};\vec k_{1},m_{1},n; 
\vec k_{2},m_{2},n|H_{\rm Cb}|\vec k_{1}',m_{1}',n;\vec
k_{2}',m_{2}',n\rangle
b_{\vec k_{1}''l_{1}}^{\dagger}
b_{\vec k_{2}''l_{2}}^{\dagger}
c_{\vec k_{1}m_{1}}^{n\dagger}
c_{\vec k_{2}m_{2}}^{n\dagger}
c_{\vec k_{2}'m_{2}'}^{n}
c_{\vec k_{1}'m_{1}'}^{n} + \mbox{H.c.}
\end{equation}
\end{widetext} 
with 
\begin{equation}
  \label{eq:1}
  \vec k_1'' + \vec k_2'' + \vec k_1 + \vec k_2 = \vec k_1' + \vec k_2' +
  \vec K
\end{equation}
where $\vec K$ stands for a vector of the reciprocal lattice.  As already
mentioned above, in addition $H_{\rm Cb}^{n\sigma}$ conserves the crystal
spin angular momentum. In bcc materials this conservation law may be
expressed by the analogous equation \cite{es3}
\begin{equation}
  \label{eq:4}
  l_1 +l_2 + m_1 + m_2 = m_1' + m_2' + 4n
\end{equation}
where $n$ is an integer. The boson operators $b_{\vec kl}^{\dagger}$ and
$b_{\vec kl}$ create and annihilate phonons with wave vector $\vec k$ and
crystal spin $l = -1, 0, +1$; the fermion operators $c_{\vec km}^{n\dagger}$
and $c_{\vec km}^n$ create and annihilate nonadiabatic Bloch states with wave
vector $\vec k$ and crystal spin $m = \pm \frac{1}{2}$.

Remember that $H_{\rm Cb}^{n\sigma}$ depends on boson operators because
during a spin-flipping process of two electrons a phonon pair is emitted (or
absorbed) in a narrow $\sigma$ band. Eq.~\gl{eq:2} shows that these flipping
processes are closely connected with scattering processes of the Bloch
electrons: a flipping of two spins effects a scattering of these electrons
within the $\vec k$ space or, vice versa, a scattering of two Bloch electrons
effects a flipping of the spins.

The mechanism of Cooper pair formation via crystal-spin-1 phonons in a narrow
$\sigma$ band is characterized by the equation of constraint (3.36) of Part
I. The physical origin of the related ``constraining forces'' may be
understood as follows. 

\begin{figure}
\includegraphics[width=.45\textwidth]{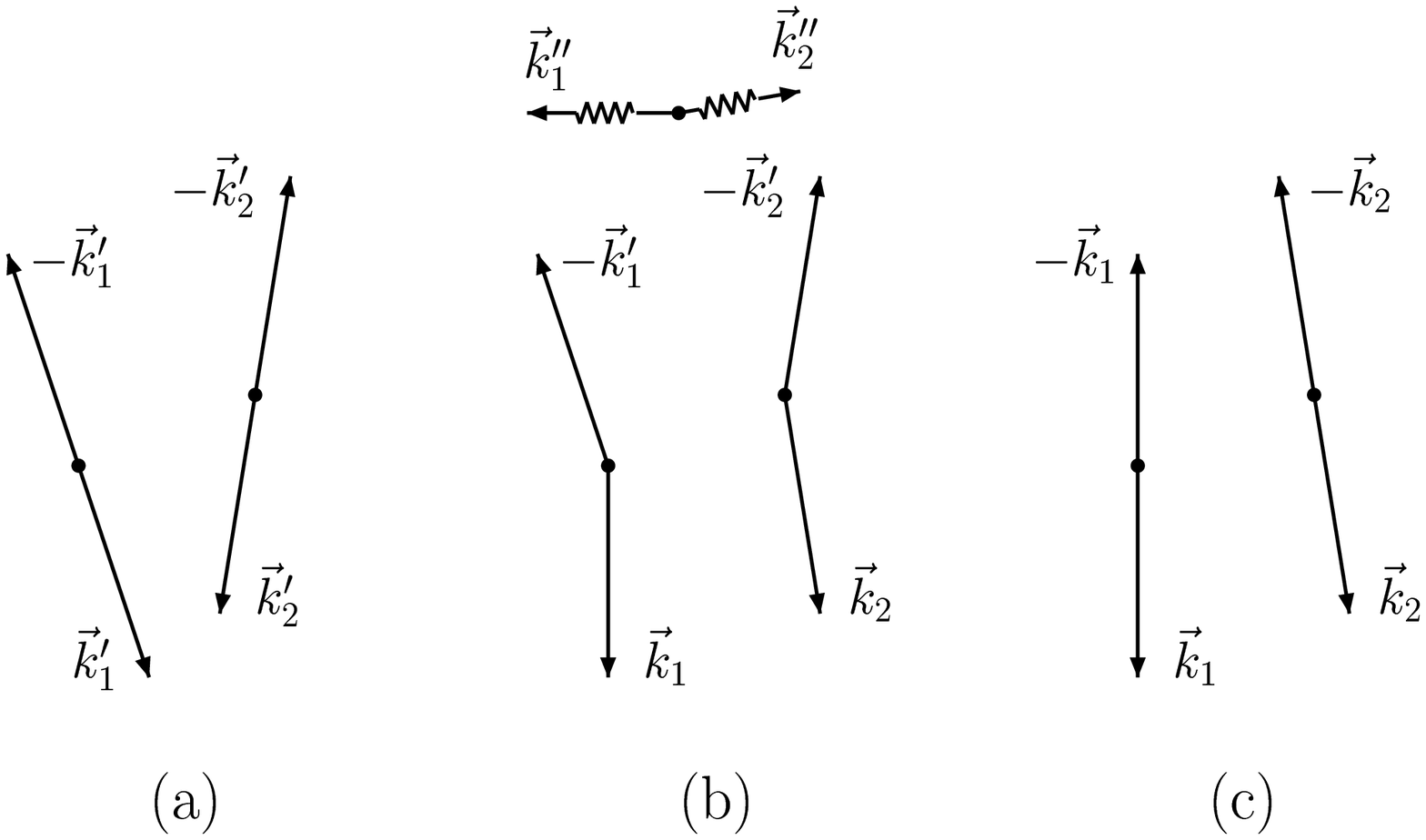}
\caption{``Spring-mounted'' Cooper pairs in a narrow $\sigma$ band:
  \\(a) Four electrons form Cooper pairs. The total crystal spin of each
  Cooper pair is zero.  \\(b) The two electrons with wave vectors $\vec k_1'$
  and $\vec k_2'$ are scattered on creating a phonon pair (with wave vectors
  $\vec k_1''$ and $\vec k_2''$).  The total crystal spin of the four
  electrons {\em and} the two phonons is zero, while neither the four
  electrons nor the two phonons alone possess a defined total crystal
  spin.\\(c) The two phonons can be completely reabsorbed only by the
  time-inverted scattering process. Hence, after the absorption of the
  phonons, the electrons again form Cooper pairs of zero total crystal spin.}
\label{figure3}
\end{figure}

Assume all the $N$ electrons (with $N$ being even) in a narrow $\sigma$ band
to form (symmetrized) Cooper pairs of the form
\begin{equation}
  \label{eq:3}
  \beta^{n\dagger}_{\vec km} = c_{\vec km}^{n\dagger}c_{-\vec
  k-m}^{n\dagger} - c_{\vec k-m}^{n\dagger}c_{-\vec
  km}^{n\dagger}; 
\end{equation}
cf. Eq.~(3.25) of Part I.  Two of these Cooper pairs are depicted in
Fig.~\ref{figure3}(a). The total crystal spin of each Cooper pair is zero.

Let be ${\cal P}^0$ the subspace of the Hilbert space spanned by the
$N$-electron states in the $\sigma$ band in which all the electrons form
Cooper pairs.  In ${\cal P}^0$ the electrons possess only $2N$ degrees of
freedom because always {\em two} electrons are characterized by four quantum
numbers $k_x$, $k_y$, $k_z$ and $m$ (cf. Section 4 of Part I).  However, the
electrons cannot stay in ${\cal P}^0$ because in reality they have $4N$
degrees of freedom. Thus, they necessarily become scattered out of this
subspace creating then a phonon pair [satisfying Eqs.~\gl{eq:1} and
\gl{eq:4}] because any scattering process is connected with a flipping of
the spins.

In Fig.~\ref{figure3}(b) the two electrons with the wave vectors $\vec k_1'$
and $\vec k_2'$ become scattered emitting then a phonon pair with the wave
vectors $\vec k_1''$ and $\vec k_2''$. The total crystal spin of the four
electrons {\em and} the two phonons still is exactly zero (since $H_{\rm
  Cb}^{n\sigma}$ conserves the crystal spin angular momentum), but the
phonons and electrons are strongly coupled in the sense that neither the two
phonons nor the four electrons {\em alone} possess a defined total crystal
spin. The strength of this coupling as well as the wave vectors of the
scattered electrons and the emitted phonons strongly depend on the special
features of the preceding spin-flipping process. As stated above, the spins
flip in a narrow $\sigma$ band in order that the localized electrons comply
with the Pauli principle. Hence, the features of the spin-flipping processes
are determined by the wave vectors of the flipping electrons as well as by
the wave vectors of all the other electrons that produce the potential in
which the scattered electrons move.

The two phonons cannot be reabsorbed spontaneously because their complete
absorption can only be the result of an appropriate motion of two other
electrons in the $\sigma$ band. Such a motion is given when these two
other electrons are {\em forced by the Pauli principle} to change their spin
directions in such a way that the phonon pair ($\vec k_1''$, $\vec k_2''$) is
completely reabsorbed.  Such a motion is performed by the time-inverted
electron pair ($-\vec k_1'$, $-\vec k_2'$): these two electrons carry out
exactly the time inverted flipping process leading to a complete absorption
of the phonon pair. Hence, after the absorption of the phonon pair the
electrons again form Cooper pairs with zero total crystal spin, as
illustrated in Fig.~\ref{figure3}(c).

This mechanism may be interpreted in terms of ``spring-mounted'' Cooper
pairs: There exist ``springs'' in the nonadiabatic system that push the
electrons back into the subspace ${\cal P}^0$. These springs become tense
when phonons get excited, and they become slacken again when these phonons
are reabsorbed.

The operator $H^0$ represents a system with these springs being rigid.  The
effect of these rigid springs may be interpreted in terms of constraining
forces that produce frozen Cooper pairs possessing half the degrees of
freedom of unpaired electrons. 

In analogy to classical physics, I suppose that only these quantum mechanical
constraining forces produced by the described ``springs'' in narrow $\sigma$
bands can produce stable Cooper pairs (with half the degrees of freedom of
unpaired electrons).  This supposition is corroborated by the calculated band
structures of those metals that experimentally prove to be superconductors;
see Part II of this paper \cite{josm}.

\begin{acknowledgements}
I thank Ernst Helmut Brandt for critical comments on the manuscript.
\end{acknowledgements}

\end{document}